# Power dissipation in spintronic devices out of thermodynamic equilibrium

Dmitri E. Nikonov[1], George I. Bourianoff, and Paolo Gargini

Abstract— Quantum limits of power dissipation in spintronic computing are estimated. A computing element composed of a single electron in a quantum dot is considered. Dynamics of its spin due to external magnetic field and interaction with adjacent dots is described via the Bloch equations. Spin relaxation due to magnetic noise from various sources is described as coupling to a reservoir. Resulting dissipation of energy is calculated and is shown to be much less than the thermal limit, ~kT per bit, if the rate of spin relaxation is much slower than the switching rate. Clues on how to engineer an energy efficient spintronic device are provided.

[1] Strategic Research, Technology & Manufacturing Group, Intel Corp., 2200 Mission College Blvd., Santa Clara, California 95052, USA

Phone 408-765-8699; fax: 408-765-2949; e-mail: dmitri.e.nikonov@intel.com

# 1    INTRODUCTION

Electronic integrated circuits have been tremendously successful in scaling towards the ever higher transistor densities and switching speeds, as was projected by the ITRS roadmap for silicon [1]. It was estimated that as this scaling approaches the nanometer scale lengths, scaling will become more difficult [2]. One element of continued scaling is successfully managing thermal dissipation. Although many elements contribute to power dissipation in modern microprocessors, one significant contribution is the basic switching element itself. Many authors e.g. [2,3] have derived the minimum energy dissipation for a switching operation (the "Shannon limit") to be:

$$E_{el} = k_B T \log 2 \,, \tag{1.1}$$

where $T$ is the ambient temperature, and $k_B$ is the Boltzmann's constant. However, the basic assumption upon which such derivations depend is that the system has time to come to thermal equilibrium between each switching event and that the equilibrium occupation of states is described by the Boltzmann distribution.

This paper extends the previous works dealing with fundamental limits of computation by considering sub-systems out of equilibrium with the thermal environment and out of local statistical equilibrium internally. These systems cannot be well described by equilibrium statistical mechanics but must be described by dynamic equations of motion, e.g. the Bloch equations. **For such non-equilibrium systems we will show that the potential energy dissipation can potentially be much less than kT per operation and that the Shannon limit does not apply.**

In order to access the degrees of freedom in such non-equilibrium systems, we must consider alternative state variables other than electronic charge. Electron spin is one such alternative state variable that receives a great deal of attention (for a recent reviews see [4,5]). A primary motivation for considering electron spin as an alternative state variable in the context of improved power efficiency is the observation that electron spins are less strongly coupled to the phonon bath than electronic charges. The weaker coupling manifests itself as spin relaxation times long compared to electron relaxation times.

The fundamental reason that the limit (1.1) applies to equilibrium systems is that all conventional computing devices, very quickly (within one switching time) achieve thermal equilibrium with the environment. This limit is avoided if the computing device satisfies the following 4 requirements:

1. The device operates far away from thermal equilibrium.

2. The device accesses energy levels spaced more closely than kT.

3. The timescale for a noise induced random event is long compared to the device switching time.

4. The thermalization time between the isolated system containing the device and the thermal reservoir is long compared to the device switching time.

Many previous authors (including Feynman) [6-9] have discussed the possibility of computing at less that the "kT limit", but have not tied the possibility directly to the thermodynamic requirements articulated above. In order to prove that a system satisfying all 4 requirements simultaneously is physically possible we analyze a spin based physical system consisting of magnetically coupled quantum dots in the presence of magnetic noise that can provide useful computational work. We calculate that the power dissipation associated with an elemental switching operation in such a system can be much less than the one determined by (1.1).

To prove this point, one needs a thorough quantum treatment of spin relaxation in the presence of random magnetic fields consistent with the phonon bath. The quantum theory of the reservoir interaction is well developed and experimentally verified for harmonic oscillators and optical transitions [10]. For the transitions between the spin states of interest here, the most relevant treatment of spin interaction with the reservoir is developed in [11]. In addition, an extensive theoretical and experimental literature exists on relaxation of free-carrier spins (see, e.g. [5]). Many fewer studies address spins of localized electrons. One of them [12] can be used as a starting point to calculate relaxation rates for our example of spins in quantum dots.

In Section 2 we describe the fundamental elements of the spintronic circuit used for the purposes of the present analysis and the procedure of computation. Section 3 lays out the quantum mechanical framework to rigorously model the interaction of a confined electron spin with the magnetic field and spins of neighboring confined electrons. Here we also state the definitions we use for the benefit of the reader not familiar with quantum mechanics of spin. Section 4 specifies the evolution of a single spin over various stages of computation. We calculate the rate of power dissipation of a spintronic device in Section 5 and demonstrate the possibility of operation with energy per bit much less than the limit (1.1). Power dissipation in a spintronic and electronic devices is compared in Section 6. Power dissipation associated with creation of magnetic fields is estimated in Section 7. Conclusions, including the possibility of non-equilibrium operation, are made in Section 8. The derivation of the Bloch equations and the rate of spin relaxation are presented in Appendix A. It follows the method of [10] and [11], though [10] only derive the general master equation, stopping short of the Bloch equations. We also avoid some of the assumptions made in [11], such as the near-field of a conducting slab and neglecting the reservoir frequencies compared to the rotating frame frequency, and explicitly show the steps of the derivation omitted in [11]. Appendix B considers in detail the mechanisms of spin relaxation and its dependence on the interaction with the rest of the circuit and reviews the results of [12]. The estimates of the noise magnetic field and the correlation time of the reservoir are made in Appendix C.

## 2 FUNDAMENTAL ELEMENTS OF SPINTRONIC CIRCUITS

We envision the elementary system for spintronic computing as a single electron confined in the conduction band within a quantum dot (Figure 1). The quantum dot might have the size of the order of 5 to 10nm such that only one confined quantum state exists. The electron may come from a dopant atom. The concentration of these atoms is chosen such that, on the average, one free electron exists for each quantum dot. The dielectric properties of a quantum dot may be engineered to ensure that not more than one electron may populate it due to the Coulomb blockade effect. The energy barrier height between the outside and the inside of the quantum dot is made much higher than the thermal energy. This would exclude the probability of escape of the electron from the quantum dot.

Thus the only remaining degree of freedom of this electron is its spin state. Projections of the spin are defined as an expectation value of the spin operators in a quantum state. The magnitude of the spin is the absolute value of the vector of spin projections. The spin quantum state can be pure, which corresponds to the maximum possible projection of spin $\hbar/2$ on a certain axis; or it can be mixed, which corresponds to a statistical sum of pure states, each existing with certain probability. The magnitude of spin can be less than $\hbar/2$ in a mixed state. A special case is a completely unpolarized state with zero projection on any axis. In any of these states the projection of spin on any axis has intrinsic quantum uncertainty.

The electron spin will be manipulated by a non-stochastic external magnetic field, which may be constant or variable in time. The external magnetic field causes precession of spin around the direction of the magnetic field without the change of the magnitude of the spin. If the external magnetic field is set to zero, the precession motion stops.

Many stochastic factors affect the spin [11,12], such as magnetic fields of the noise currents in wires, spin-orbit coupling to the vibrations of the crystal (phonons), vacuum fluctuations of the electromagnetic field, etc. All these factors contribute to spin relaxation. They are simultaneously manifested as both damping of the spin magnitude and the magnetic noise [11]. In other words, unlike an external magnetic field, a stochastic factor can change the magnitude of spin. Damping is characterized via the spin relaxation rate. It is also linked to the intensity of magnetic noise (its auto-correlation function) via the fluctuation-dissipation theorem. Spin relaxation happens due to interaction of the system with numerous degrees of freedom ("reservoir") which are typically in a state of thermodynamic equilibrium defined by the ambient temperature $T$. The considered single spin is not necessarily in thermal equilibrium with the reservoir. See Section 3 for the mathematical treatment of the spin-reservoir interaction.

The interaction of the single spin with the reservoir drives the magnitude of the spin to its equilibrium value which is determined by the non-stochastic magnetic field. Its only projection in equilibrium is along the direction of this field. This part of the evolution happens as damping, typically with exponential dependence on time, and not as precession. The effect of the stochastic field from the reservoir on the single spin is quite different from the effect of the applied, non-stochastic field due to the very different time scale of its variation. The fluctuations of the magnetic noise from the reservoir are happening at a characteristic frequency (determined by the cut-off frequency of the reservoir) which is much larger than any characteristic frequencies of the spin subsystem. The inverse of the cut-off frequency determines the correlation time of the reservoir. Thus the fluctuating magnetic field averages out as far as the projections of the spin are concerned. The above description of spin dynamics is applicable to values averaged over times much longer than the correlation time of the reservoir.

It is also well-known [12] that the spin relaxation rate strongly depends on the magnitude of the non-stochastic magnetic field. For example, the relaxation rate due to interaction with phonons varies approximately as the fifth power of the magnetic field. As we see in Section 3, the relaxation rate will be much slower than the precession rate if the magnetic field is small, and will be much faster than the precession rate, if the magnetic field is large. Thus the dynamics of spin is always represented as damped precession. However for small magnetic fields, precession dominates, and for large magnetic fields, damping dominates.

With this understanding of spin dynamics we propose a procedure for spintronics computing that will operate at less than the Shannon limit (1.1). We represent the binary information by the projection of a single spin on the z-axis.

a) The spin is prepared in a state with large projection of spin (comparable but smaller than $\hbar/2$) on the x-axis. This is accomplished by applying a strong external magnetic field along the x-axis (Figure 2a). To maintain agreement with earlier works we will call this field "clocking" [13]. Since the field is strong, the relaxation rate in its presence is shorter than a switching cycle. Therefore the spin is quickly damped to equilibrium with the field. After that the clocking field is removed, but the magnitude of the spin remains.

b) The spin is prepared in its initial state with either positive or negative projection on the z-axis. This is done by applying a weaker "setting" magnetic field along the y-axis for a definite duration (Figure 2b). This field causes precession of spin by 90 degrees from the x-axis to the z-axis, as explained above.

c) Computing is envisioned as the switching of the spin state depending on the input magnetic field. It is performed by applying a similarly weaker "computing" magnetic field along the y-axis for a definite duration (Figure 2c). This field causes precession of spin by 180 degrees. The direction of spin remains close to the z-axis;

d) One can read-out the spin after any number of switching cycles. For theoretical simplicity we envision the read-out with the magnetic force microscope (MFM), Figure 2d.

e) During computing, the magnitude of the spin is being damped by spin relaxation. Since the external magnetic field is relatively small in stages b)-d), the relaxation time is much longer than the switching time. Therefore it is possible to perform many (ideally thousands) of switching cycles before the spin magnitude is degraded, see Figure 2e. After that a refresh of the spin magnitude is needed with the aid of the "clocking" field. The spin output states need to be read-out just before clocking and stored in memory. They are then needed as inputs in the next clocking cycle. We return to stage a).

A more practical variation of this spintronics element is one of multiple quantum dots with a single electron in each, see Figure 3. In this case the computing is done via interaction of spins of single electrons with the spins of neighboring confined electrons (due to the exchange part of the Coulomb force), and not with the magnetic fields acting individually on each single electron. Initial setting of spins is performed by injecting spin current from ferromagnetic contacts attached to a few input quantum dots. The dots are formed into geometry patterns which implement logic gates (AND, OR, etc.) [14] or majority gates [15]. The read-out is performed by passing current from a ferromagnetic electrodes attached to a few output quantum dots. The current depends on the orientation of spin to a magnetization of a fixed ferromagnetic layer, as a version of the giant magneto-resistance effect [14].

### 3    QUANTUM MECHANICS OF SPIN EVOLUTION

The Hilbert space of the spin=½ has the dimension of 2. The basis of this space may be chosen as the states with the spin projection of $+\hbar/2$ and $-\hbar/2$ on any chosen axis, e.g. z-axis. Further we will call them states with spin up ("u") and spin down ("d"), respectively. Any pure or mixed quantum state can be identified by the density matrix in the above basis

$$\rho = \begin{pmatrix} \rho_{uu} & \rho_{ud} \\ \rho_{du} & \rho_{dd} \end{pmatrix}. \tag{3.1}$$

All physical quantities correspond to operator matrices

$$\hat{O} = \begin{pmatrix} O_{uu} & O_{ud} \\ O_{du} & O_{dd} \end{pmatrix}. \tag{3.2}$$

For example, the operator of the spin vector is

$$\mathbf{s} = \frac{\hbar}{2}\boldsymbol{\sigma}, \tag{3.3}$$

where the Pauli matrices correspond to projections the spin on each axis

$$\sigma_x = \begin{pmatrix} 0 & 1 \\ 1 & 0 \end{pmatrix}, \ \sigma_y = \begin{pmatrix} 0 & -i \\ i & 0 \end{pmatrix}, \ \sigma_z = \begin{pmatrix} 1 & 0 \\ 0 & -1 \end{pmatrix}. \tag{3.4}$$

The expectation value of any physical quantity in a quantum state is determined as

$$\langle O \rangle = Tr\left[\hat{O}\rho\right]. \tag{3.5}$$

This results in the following relation between the elements of the density matrix and the projections of spin

$$\langle \sigma_x \rangle = \rho_{ud} + \rho_{du}, \tag{3.6}$$

$$\langle \sigma_y \rangle = i\rho_{ud} - i\rho_{du}, \tag{3.7}$$

$$\langle \sigma_z \rangle = \rho_{uu} - \rho_{dd}. \tag{3.8}$$

The elements $\rho_{uu}$ and $\rho_{dd}$ are probabilities of the spin pointing up and down the "z" axis, respectively. Their sum must be equal to the number of particles, thus

$$p_{uu} + p_{dd} = 1, \tag{3.9}$$

The real and imaginary parts of off-diagonal elements are related to the projection of the spin on the "x" and "y" axes, respectively, and must obey

$$\rho_{ud} = \rho_{du}^*, \tag{3.10}$$

where the star designates a complex conjugate. In other words the density matrix is Hermitean. Below are a few special cases of the density matrix corresponding to: spin in the positive "x" direction, spin in the positive "z" direction, and unpolarized spins

$$\rho(s_x = -\hbar/2) = \begin{pmatrix} 1/2 & -1/2 \\ -1/2 & 1/2 \end{pmatrix}, \rho(s_z = \hbar/2) = \begin{pmatrix} 1 & 0 \\ 0 & 0 \end{pmatrix},$$

$$\rho(s = 0) = \begin{pmatrix} 1/2 & 0 \\ 0 & 1/2 \end{pmatrix}. \tag{3.11}$$

An interesting property of the spin-1/2 is that the operators of squares of spin projections are proportional to unity matrices

$$\sigma_x^2 = \sigma_y^2 = \sigma_z^2 = \begin{pmatrix} 1 & 0 \\ 0 & 1 \end{pmatrix}. \tag{3.12}$$

This results in the variances of spin projections being

$$(\delta s_x)^2 \equiv \langle s_x^2 \rangle - \langle s_x \rangle^2 = \frac{\hbar^2}{4} - \langle s_x \rangle^2. \tag{3.13}$$

For example in the completely unpolarized spin state, the variances of all projections are $\hbar^2/4$. In the state with the spin $+\hbar/2$ along z-axis, the variance of z-projection is zero, but the variances of the other two projections are still $\hbar^2/4$. Also any power of the spin operator can be expressed via a constant or the first power of the spin operator. This means that the expectation values of projections of the spin contain the complete information about the state of the spin. Note that the variance here has the intrinsic quantum character. Unlike in classical statistical physics, it is impossible to pick a member of the ensemble with projections different than the expectation values and then follow its evolution, see Figure 4.

The operator corresponding to energy is the Hamiltonian $H$. It determines the evolution of any expectation value according to the Heisenberg's equation

$$\frac{d\langle O \rangle}{dt} = \frac{i}{\hbar}\langle [H, \hat{O}] \rangle. \tag{3.14}$$

We separate the magnetic field acting on the spin into two parts: the non-stochastic magnetic fields $\mathbf{B}$ and the stochastic effective magnetic field due to interaction with the reservoir. The non-stochastic field includes in our case the clocking, setting and computing magnetic fields $\mathbf{B} = \mathbf{B}_c(t) + \mathbf{B}_{set}(t) + \mathbf{B}_0(t)$.

The Hamiltonian corresponding to the spin degrees of freedom of the single electron under consideration is thus

$$H = H_B + H_{ss} + H_{res}. \tag{3.15}$$

The first term in (3.15) describes the interaction of the spin with external non-stochastic magnetic field $\mathbf{B}$:

$$H_B = -G\mathbf{B} \cdot \mathbf{s}. \tag{3.16}$$

The gyromagnetic ratio

$$G = \frac{-ge}{2m_e} = \frac{-g\mu_B}{\hbar}, \tag{3.17}$$

where $e$ is the absolute value of the electron charge, $m_e$ is the mass of a free electron, $\mu_B$ is the Bohr's magneton, and $g$ is the Lande g-factor.

The second term in (3.15) describes interaction with the spins of electrons in neighboring quantum dots (they are labeled by the index "j")

$$H_{ss} = -\sum_j A_j \mathbf{s} \cdot \mathbf{s}_j, \tag{3.18}$$

where $A_j$ is the coupling constant which depends on the overlap of the wavefunctions of the states in the quantum dots, dielectric constant, etc.

The remaining third term in (3.15) describes interaction with a reservoir which can represent magnetic noise due to phonons, due to fluctuating current in wires, or due to fluctuations of the electromagnetic field [10]

$$H_{res} = \sum_n \frac{1}{2}\left\{ g_n^* a_n + g_n a_n^+, \mathbf{n} \cdot \mathbf{s} \right\}. \tag{3.19}$$

The summations is performed over the modes of the reservoir, $\mathbf{n}$ is a unit vector corresponding to a reservoir mode, the coupling constants are $g_n$, creation and annihilation operators for the reservoir mode are $a_n^+, a_n$, and $\{,\}$ designates the anti-commutator.

The non-stochastic terms (3.16) and (3.18) in the Hamiltonian, referred to as the "system", result in the precession of the expectation values of projections of the spin with the angular velocity (as we will verify further in (3.24))

$$\mathbf{\Omega} = G\mathbf{B} + A_j \langle \mathbf{s}_j \rangle. \tag{3.20}$$

In the special case when the z-axis is chosen along the direction of $\mathbf{\Omega}$, the non stochastic part of the Hamiltonian assumes the diagonal form

$$H_B + H_{ss} = \frac{\hbar\Omega}{2}\begin{pmatrix} -1 & 0 \\ 0 & 1 \end{pmatrix}. \tag{3.21}$$

This expresses the Zeeman splitting of the spin energy levels. The state of thermodynamic equilibrium under the influence of the field (3.20) is

$$\rho = \frac{1}{Z}\exp\left(-\frac{H_B + H_{ss}}{k_B T}\right), \tag{3.22}$$

where, $Z$ is the normalization factor ("statistical sum"). This translates into the equilibrium projections of the spin

$$\mathbf{s}_{eq} = \frac{\hbar}{2}\tanh\left(\frac{\hbar\Omega}{2k_B T}\right)\hat{\mathbf{\Omega}}, \tag{3.23}$$

where $\hat{\mathbf{\Omega}}$ is the unit vector in the direction of $\mathbf{\Omega}$. This is the expectation value that the spin tends to due to spin relaxation after a sufficiently long time.

Even though the Hamiltonian (3.19) is proportional to the first power of spin, similarly to (3.16) and (3.18), its effect on the evolution of the spin is different. The reason for that is that the reservoir has numerous modes which form a quasi-continuous spectrum which stretches up to a certain maximum frequency of the reservoir modes (the "cut-off frequency" $\omega_c$, see Appendix B). The reservoir modes are considered to be in thermal equilibrium at the ambient temperature $T$. The phase and occupation number each modes are random. Therefore the angular velocity resulting from the interaction with the reservoir is stochastic, i.e. it experiences random variations on the time scale of $\tau_{corr} \sim \omega_c^{-1}$ (see Appendix C). For most of the practical cases of interest this time scale is much shorter than the characteristic times of evolution of the spin. This stochastic part of precession manifests itself as damping of the expectation value of the spin and change of the variance of the projections (see Appendix A for derivation).

As a result, the total evolution of the spin is described by the Bloch equation and is composed of two parts: the precession with angular velocity (3.20) according to (3.14) and the spin relaxation due to interaction with the reservoir (Appendix A)

$$\dot{\mathbf{s}} = -\mathbf{\Omega} \times \mathbf{s} - \mathbf{\gamma}\left(\mathbf{s} - \mathbf{s}_{eq}\right). \tag{3.24}$$

For the Bloch equation applied to free electrons see [5]. The equation is formulated for the vector of expectation values of the projections $\mathbf{s}$ (we will drop the angle brackets for them henceforth). As we argued before, this provides a complete description of spin dynamics. In the Bloch equation, $\mathbf{\gamma}$ is the relaxation matrix (A.37). Both longitudinal and transverse spin relaxation rates depend on the spin splitting $\Omega$ (see Appendix A). We will be mostly interested in the longitudinal spin relaxation rate $\gamma_l(\Omega)$. It is determined by the dependence of the reservoir mode density and of coupling constants on the frequency of the modes. Only modes with frequencies, which are resonant to $\Omega$, contribute strongly to spin relaxation.

Specific examples of spin relaxation due to phonon interaction, noise currents in wires, and fluctuations of the electromagnetic field are considered in Appendix B. For present purposes it is sufficient to know that the relaxation rate (see Figure 5) may be a steep function of $\Omega$ and it has a high cut-off frequency.

Now armed with this knowledge of the spin dynamics, we can calculate the change of spin projections and energy over the clocking cycle.

## 4 Spin evolution and time scales

The Bloch equation (3.24) enables us to calculate the evolution of spin over all parts of the clocking cycle of Section 2.

TABLE I
FREQUENCY SCALES IN COMPUTATION

| Quantity | Frequency | Frequency [1/s] | Corresp. energy | Energy [eV] |
|---|---|---|---|---|
| Relaxation rate during computing | $\gamma_l(\Omega_0)$ | 2e7 | $\hbar\gamma_l(\Omega_0)$ | 1.3e-8 |
| Angular velocity during computing | $\Omega_0$ | 1.5e12 | $\hbar\Omega_0$ | 1e-3 |
| Thermal energy | $k_B T / \hbar$ | 4e13 | $k_B T$ | 2.6e-2 |
| Angular velocity during clocking | $\Omega_c$ | 7.9e13 | $\hbar\Omega_c$ | 5.2e-2 |
| Cutoff frequency | $\omega_c$ | 1.4e14 | $\hbar\omega_c$ | 7.5e-2 |
| Relaxation rate during clocking | $\gamma_l(\Omega_c)$ | 1.2e15 | $\hbar\gamma_l(\Omega_c)$ | 0.78 |

We will specify numerical values for the angular velocities and relaxation rates (see Table I) for illustrative purposes only. The conclusions still apply for the same relative relation between time scales. The relaxation rates are represented in Figure 5. The ambient temperature is taken to be the room temperature, 300K. The material parameters are as described in Appendix B.

a) For the clocking stage we apply a magnetic field $\mathbf{B}_c$ along the x-axis, which produces the splitting

$$\Omega_c = GB_c. \tag{4.1}$$

This splitting corresponds to a very fast relaxation rate $\gamma_l(\Omega_c)$. Thus the thermal equilibrium will be reached in a very short time after the clocking field is switched on

$$\tau_{eq} \sim \left[\gamma_l(\Omega_c)\right]^{-1} \sim 1\text{fs} .$$
(4.2)

much shorter than any other time scale, see Figure 6. The spin state in equilibrium is

$$\mathbf{s}_c = \frac{\hbar}{2}\tanh\left(\frac{\hbar G B_c}{2k_B T}\right)\hat{\mathbf{x}} .$$
(4.3)

For the numerical example we choose the splitting to be

$$\Omega_c = G B_c \approx 2k_B T .$$
(4.4)

b) For both setting and computing, a magnetic field $\mathbf{B}_0$ is applied such that its strength produces splitting

$$\Omega_0 = G B_0 .$$
(4.5)

as in Table I. The relaxation at such splitting is very slow, so the magnitude of the spin vector approximately remains the same and precesses with the angular velocity (4.5) around the direction of the magnetic field. For setting, the field $\mathbf{B}_0$ is along the y-axis, and it takes the time

$$\tau_{set} = \frac{\pi}{2\Omega_0} .$$
(4.6)

to rotate the spin to the x-axis.

c) Computing is done by either applying or not applying a field $\mathbf{B}_0$ along the y-axis in each switching cycle of duration

$$\tau_{sw} = \frac{\pi}{\Omega_0} \approx 2\text{ps} .$$
(4.7)

d) In order to measure a spin state, a field of a magnetic force microscope (MFM) is applied vertically, along the z-axis. In order to make the measurement in the time of one switching cycle, a spin splitting of at least $\Omega_0$ is required, according to the time-energy uncertainty relation (or more rigorously the Margolis-Levitin theorem [16]). It means that the field must be at least $B_0$. In the measurement, the direction of spin does not change; the spin is just determined to have either negative of positive projection in the z-axis.

e) After many switching cycles, the magnitude of spin will decrease, and by the end of the clocking cycle, time $\tau_c$ it will be

$$s_{end} = s_c \exp\left(-\gamma_l(\Omega_0)\tau_c\right) .$$
(4.8)

In order to get an accurate read-out of the spin state, this magnitude must be comparable to $s_+$, which translates in the condition that

$$\gamma_l(\Omega_0)\tau_c \leq 1. \tag{4.9}$$

For the numerical example we take the clocking cycle time such that the spin magnitude decreases by ~25%

$$\tau_c = 0.3 / \gamma_l(\Omega_0) \approx 15\text{ns}. \tag{4.10}$$

Thus we see that many thousands of switching cycles can be performed before the spin magnitude needs to be refreshed by clocking.

## 5    POWER DISSIPATION IN SPINTRONIC COMPUTATION

Once we know the evolution of spin over the clocking cycle, we are in a position to calculate the dissipated power. For this we will track the changes in the energy of the spin system

$$E(t) = \left\langle H_B + H_{ss} \right\rangle. \tag{5.1}$$

Note that the energy (3.19) of interaction between the spin system and the reservoir is not included in this equation since it averages to zero. The energy (5.1) will increase as the work will be performed by external fields on the system and it will decrease as the energy is dissipated into the environment due to interaction with the reservoir. In terms of the spin projections, the energy is recast as

$$E(t) = -\mathbf{\Omega} \cdot \mathbf{s}. \tag{5.2}$$

a) At the beginning of the clocking state, the spin is along the z-axis, while the clocking field is along the x-axis, therefore

$$E(0) = 0. \tag{5.3}$$

As the clocking field is switched on, the relaxation by the reservoir causes the spin to align to the clocking field and thus decreases its energy

$$E(\tau_{eq}) = -\Omega_c s_c. \tag{5.4}$$

Then as the clocking field is switched off, it performs work on the spin and the energy of the spin returns to zero.

$$E(2\tau_{eq}) = 0. \tag{5.5}$$

In other words, the field replenishes the energy dissipated in a clocking event

$$\Delta E_{clock} = \Omega_c s_c = \frac{\hbar\Omega_c}{2}\tanh\left(\frac{\hbar\Omega_c}{2k_B T}\right). \tag{5.6}$$

b, c) In setting input values and in computing, the spin rotates around y-axis, the direction of the magnetic field. But only the small y-projection determines the change of energy. To calculate it one needs to consider all projections of the Bloch equation

$$\dot{s}_y = -\gamma_l(s_y - s_0),\tag{5.7}$$

$$\dot{s}_z = \Omega_0 s_x - \gamma_t s_z,\tag{5.8}$$

$$\dot{s}_x = -\Omega_0 s_z - \gamma_t s_x,\tag{5.9}$$

and here

$$s_0 = \frac{\hbar}{2}\tanh\left(\frac{\hbar\Omega_0}{2k_B T}\right).\tag{5.10}$$

Over many switching cycles, see Figure 7, the average of $s_y$ is zero, therefore the change of energy

$$\frac{dE}{dt} = -\Omega_0 \gamma_l(\Omega_0)s_0.\tag{5.11}$$

On the average, the dissipation per switching cycle is

$$\Delta E_{sw} = t_{sw}\gamma_l(\Omega_0)\Omega_0 s_0$$
$$= \pi\gamma_l(\Omega_0)\frac{\hbar}{2}\tanh\left(\frac{\hbar\Omega_0}{2k_B T}\right).\tag{5.12}$$

d) In the read-out, the field is applied along the z-axis, and the

of the crystal field and the energy in this field which is later dissipated to the reservoir is

$$\Delta E_{read} = \Omega_0 s_c = \frac{\hbar\Omega_0}{2}\tanh\left(\frac{\hbar\Omega_c}{2k_B T}\right).\tag{5.13}$$

Now collecting the contributions to the power dissipation (5.6), (5.12) and (5.13), and normalizing them to one switching cycle, results in

$$\Delta E_{tot} = \Delta E_{clock}\frac{\tau_{sw}}{\tau_c} + \Delta E_{sw} + \Delta E_{read} / N_{read}.\tag{5.14}$$

For the numerical example here we assume that read-out is performed every $N_{read} \sim 1000$ switching cycles. The number of read-off events within the clocking cycle will depend on the application and may be much smaller. Substituting the condition (4.4) and using the fact that $\hbar\Omega_0 \ll k_B T$, one obtains

$$\Delta E_{tot} = k_B T \tanh(1) \frac{\tau_{sw}}{\tau_c} + \frac{\pi \hbar \Omega_0 \hbar \gamma_l (\Omega_0)}{4 k_B T}$$
$$+ \frac{\hbar \Omega_0}{2 N_{read}} \tanh(1) \qquad . \qquad (5.15)$$

The dominant contribution to the relaxation is represented by the first term in (5.15) corresponding to clocking. We see that it is smaller than the limit (1.1) by the factor of switching time divided by the clocking time. This ratio is enabled by the much slower spin relaxation time than the switching time. For the numerical values in Table I, this reduces to

$$\Delta E_{tot} = k_B T \left( 10^{-4} + 2.3 \cdot 10^{-9} + 1.5 \cdot 10^{-5} \right). \qquad (5.16)$$

The numerical estimates are made here is only an example, but as long as the spin relaxation rate remains slower than the switching rate, the conclusion remains: together the sources of energy dissipation of energy (5.6), (5.12) and (5.13) are less than the result for electronic circuits (1.1).

## 6    COMPARISON WITH ELECTRONIC COMPUTATION

It is also necessary to point out the differences which led to fundamentally different limits of power dissipation in electronic (1.1) and spintronic (5.16) computing. For the paradigm of the electronic computing we will take the model of [2].

TABLE II
ATTRIBUTES OF COMPUTING

|  | Electronic computing | Spintronic computing |
|---|---|---|
| State | Electron to the right or to the left of the barrier | Spin has positive or negative projection |
| Write | Electron injection from a wire | Magnetic field orients spin |
| Switching | Electron flow when barrier lowered | Precession in magnetic field, no barrier |
| Read | Electron removal by a reservoir (wire) | Spin direction sensed by magnetic field |
| Clocking (refresh) | Every switching cycle | Once in many cycles |
| Relaxation rate | Faster than switching | Slower than switching |

These differences are summarized in Table II. In a typical transistor an electron is injected from a wire, which is a reservoir in thermal equilibrium. It interacts strongly with the phonons of the material, which keeps him in thermal equilibrium. A barrier with the energy height much larger than (1.1) is required to maintain the logical state. Finally the electron is removed into another wore-reservoir. The example of spintronic computing presented here insulates the spin from thermal reservoir for the duration of all switching cycles.

The ultimate cause of the different limit is that our spintronic element was kept out of thermal equilibrium with the environment.

Let us consider what would happen if the relaxation rate was faster or comparable to the switching rate. Then clocking will not be necessary. The computing field will be applied along the z-axis. Its magnitude must be comparable to the former clocking field $\Omega_c$. The value of spin in equilibrium with this field will still be $s_c$. If the computing field does not change, neither does the spin state. If it does change, the switching will amount to damping of the spin projections to a new equilibrium value. This operation will be similar to one described in [14]. Then the change of energy in one switching cycle is

$$\Delta E = -\Omega_c s_c - \Omega_c s_c , \qquad (6.1)$$

and the average energy dissipation per bit (with a factor ½ for the probability that the computing field remains the same)

$$\Delta E_{eq} = \frac{\hbar \Omega_c}{2} \tanh\left(\frac{\hbar \Omega_c}{2 k_B T}\right) \approx k_B T \tanh(1) . \qquad (6.3)$$

Therefore in the case of spintronic computing in thermal equilibrium we would arrive to a limit very similar to the electronic limit (1.1).

## 7 DISSIPATION OF MAGNETIC FIELD ENERGY

Another source of power dissipation is associated with creating magnetic fields. When the clocking or computing magnetic field is on, energy is stored in the magnetic fields produced by wire coils. When the field is switched off, a part of this energy, determined by the Q-factor of the circuit is dissipated via the resistance of the wires. Therefore the corresponding power dissipation is

$$\Delta E = \frac{B^2}{2\mu_0} \frac{V}{Q} , \qquad (7.1)$$

where $V$ is the volume of the magnetic field per spintronic device. It is determined by pitch of the devices ($a = 10\,\mathrm{nm}$). We envision that the clocking field is global, i.e. the corresponding coil wraps around the chip, while the computing coils wraps around an individual quantum dot. Thus the height of the clocking loop is ~10 larger than the pitch. Then dissipated energy in clocking is

$$E_{mag,clock} = \frac{B_c{}^2}{2\mu_0} \frac{10a^3}{Q} \frac{\tau_{sw}}{\tau_c} , \qquad (7.2)$$

and in computing

$$E_{mag,sw} = \frac{B_0{}^2}{2\mu_0} \frac{a^3}{Q} , \qquad (7.3)$$

The magnetic field corresponding to clocking (4.4) and computing (4.5) depend on the value of g-factor one can achieve.

( $g = 200$ in our case) and are estimated (C.3) to be

$$B_c = 4.46 \text{T} , \quad B_0 = 85 m\text{T} ,$$ (7.4)

The resulting estimate of energy (Q=1000)

$$E_{mag,clock} = 4.2 \cdot 10^{-25} \text{J} , \quad E_{mag,sw} = 2.8 \cdot 10^{-25} \text{J} ,$$ (7.5)

which happens to be still smaller than the thermal energy (1.1). Also note that in the scheme of Figure 3, the second source of magnetic dissipation would be absent, since spin switching is performed by interaction of neighboring dots rather than the magnetic field of a coil.

Overall, the sources of dissipation (7.2) and (7.3) are not intrinsic to the spintronic device. They are counterparts of interconnect power losses in electronic devices and should be compared as such.

## 8   CONCLUSION

We have considered a simple scheme of spin computation. It involved single electrons in each of the array of quantum dots. We analyzed the mechanisms of relaxation, and showed that its rate can be made slow during switching, and can be drastically increased during clocking. We also estimated power dissipated in such relaxation. The estimates show that on average energy much less than $k_B T$ per switching can be dissipated, if the relaxation rate is much slower than the switching rate.

The above considerations provide some clues on how a spintronic device, operating out of thermal equilibrium, could be implemented. As shown in Section 4, thousands of spin switching cycles, each of them a few picoseconds in duration, could be performed within a refresh clocking cycle occurring once every tens of nanoseconds. The ratio between operation time and refresh time of this spintronic device would somewhat resemble some of the functional aspects of DRAM devices which operate out of electrical equilibrium.  In this case, after a write cycle, some capacitors store electrical charge while others are fully depleted of charge. However, electrical charge is continuously thermally generated in depleted capacitors until it reaches the equilibrium value and all the information is lost. Refresh cycles of the order of milliseconds are necessary to preserve the charge, or absence of charge, related information stored in each of the capacitors while the presence or absence of charge under individual capacitor can be read in tens of nanoseconds.

The authors believe that spintronic devices, with features in the tens of nanometers and operated in accordance with the principles outlined in this publication, have the potential of performing computation at higher speed than nowadays corresponding CMOS devices while consuming only a fraction of the power since individual spin switching operations can dissipate much  less that $k_B T$ .

The authors are grateful to V. V. Zhirnov, J. A. Sidles, and V. V. Dodonov for extensive and helpful discussions.

## 9    APPENDIX A. DERIVATION OF BLOCH EQUATIONS.

In this appendix we derive the Bloch equations for the spin dynamics and relaxation starting from the Heisenberg equations for the interaction of a spin with the thermal reservoir. This derivation is based on the method of [10] and closely follows the approach of [11]. However we avoid some of the assumptions made in [11], such as the near-field of a conducting slab and neglecting the reservoir frequencies compared to the rotating frame frequency, and explicitly show the steps of the derivation omitted in [11] and put it in the form convenient for the presentation in the main part of our article.

We start with a general Hamiltonian of a quantum system $H_{sys}$ coupled with the reservoir of harmonic oscillators with frequencies $\omega_n$ and with creation and annihilation operators $a_n^+, a_n$:

$$
\begin{aligned}
H = H_{sys} + \sum_n \frac{\hbar \omega_n}{2} \left( a_n^+ a_n + a_n a_n^+ \right) \\
+ \sum_n \frac{\hbar}{2} \left\{ g_n^* a_n + g_n a_n^+, X \right\}
\end{aligned}
\qquad\qquad \text{(A.1)}
$$

The coupling constants $g_n$ are associated with a certain system operator $X$; $\{,\}$ is the anti-commutator. Then, according to [10], the evolution of any system operator $Y$ is determined by a solution of the Heisenberg equations

$$
\frac{dY}{dt} = \frac{i}{\hbar} \left[ H, Y \right]. \qquad\qquad \text{(A.2)}
$$

After solving for the variables of the reservoir oscillators and substituting them into (A.2), one obtains the quantum Langevin equation [10]

$$
\begin{aligned}
\frac{dY(t)}{dt} = \frac{i}{\hbar} \left[ H_{sys}, Y(t) \right] + \frac{i}{2\hbar} \{ f(t) X(0) + \\
\int_0^t f(t-t') \dot{X}(t') dt' - \xi(t), \left[ X, Y \right] \}
\end{aligned}
\qquad\qquad \text{(A.3)}
$$

where the random Langevin force is

$$
\begin{aligned}
\xi(t) = -\hbar \sum_n \left( g_n^* a_n(0) e^{-i\omega_n t} + g_n a_n^+(0) e^{i\omega_n t} \right) \\
= -\hbar \sum_n \xi_n(t)
\end{aligned}
\qquad\qquad \text{(A.4)}
$$

and the reservoir coupling function is

$$
f(t) = 2\hbar \sum_n |g_n|^2 \cos(\omega_n t) / \omega_n. \qquad\qquad \text{(A.5)}
$$

Now we specify the above formalism to the case of a spin system with the Hamiltonian describing interaction with a magnetic field or another spin

$$H_{sys} = -\mathbf{\Omega} \cdot \mathbf{s} .$$ (A.6)

Where $\mathbf{\Omega}$ has the meaning of the rotation frequency due to this coherent interaction (called so to distinguish it from the interaction with the reservoir.

In this case the operators are ($\mathbf{n}$ is a unit vector corresponding to a reservoir mode)

$$X = \mathbf{s} \cdot \mathbf{n} / \hbar , \qquad Y = \mathbf{s}$$ (A.7)

Straightforward application of (A.3) to (A.7) results in the dynamic equation for the spin

$$\dot{\mathbf{s}} = -\mathbf{\Omega} \times \mathbf{s} - \frac{1}{2} \left( \mathbf{\Omega}_t \times \mathbf{s} - \mathbf{s} \times \mathbf{\Omega}_t \right)$$
$$-\frac{1}{2} \left( \mathbf{C} \cdot \mathbf{s} \times \mathbf{s} - \mathbf{s} \times \mathbf{C} \cdot \mathbf{s} \right) \qquad .$$ (A.8)
$$+\frac{1}{2} \int_0^t (\mathbf{G}(t-t') \cdot \dot{\mathbf{s}}(t') \times \mathbf{s}(t) \\ -\mathbf{s}(t) \times \mathbf{G}(t-t') \cdot \dot{\mathbf{s}}(t')) dt'$$

The action of the reservoir is expressed via the stochastic rotation frequency

$$\mathbf{\Omega}_t = \sum_n \xi_n(t) \mathbf{n} .$$ (A.9)

and the reservoir memory tensors are

$$\mathbf{C} = \sum_n \frac{2\hbar \left| g_n^2 \right|}{\omega_n} \mathbf{n} \otimes \mathbf{n} .$$ (A.10)

$$\mathbf{G}(t) = \sum_n \frac{2\hbar \left| g_n^2 \right| \cos(\omega_n t)}{\omega_n} \mathbf{n} \otimes \mathbf{n} .$$ (A.11)

The term containing $\mathbf{C}$ turns to zero for spin ½, as noted in [11].

Further we need to perform the summations over the reservoir variables in the above formulas and to substitute these variables by a few reservoir parameters such as the relaxation rate or temperature of the reservoir. In order to correctly calculate the relaxation rates, one first needs to transform the equations to the "rotating frame", or in other words, to make a variable substitution which eliminates the first right-hand-side term in (A.8) corresponding to the interactions other than the reservoir, "coherent evolution".

The transform, in agreement with [11], is given by the orthogonal rotation matrix $\mathbf{R}(t)$ with the property

$$\mathbf{R}^T(t) = \mathbf{R}^{-1}(t) = \mathbf{R}(-t)$$ (A.12)

as follows

$$\mathbf{s}^r = \mathbf{R}\mathbf{s}, \qquad \mathbf{\Omega}_t^{\ r} = \mathbf{R}\mathbf{\Omega}_t,$$ (A.13)

$$\mathbf{G}^r(t-t') = \mathbf{R}(t)\mathbf{G}^r(t-t')\mathbf{R}^T(t')$$ (A.14)

The rotation matrix must satisfy the equation

$$\dot{\mathbf{R}} = \mathbf{\Omega} \times \mathbf{R}$$ (A.15)

which results in invariance of the coherent field under this transform – if rotation is around $\mathbf{\Omega}$, it is not affected by the rotation

$$\mathbf{\Omega} = \mathbf{R}\mathbf{\Omega}.$$ (A.16)

The transformed spin evolves according to

$$\dot{\mathbf{s}}^r = \mathbf{R}\dot{\mathbf{s}} + \mathbf{\Omega} \times \mathbf{R} \cdot \mathbf{s}.$$ (A.17)

Using the explicit form of the rotation matrix in the coordinates where $\mathbf{\Omega}$ is along the z-axis

$$\mathbf{R} = \begin{pmatrix} \cos(\Omega t) & -\sin(\Omega t) & 0 \\ \sin(\Omega t) & \cos(\Omega t) & 0 \\ 0 & 0 & 1 \end{pmatrix}.$$ (A.18)

One can show that in these coordinates and without loss of generality

$$\mathbf{R}^{-1}\mathbf{\Omega} \times \mathbf{R} \cdot \mathbf{s} = \mathbf{\Omega} \times \mathbf{s}.$$ (A.19)

i.e. the coherent evolution terms in fact cancel. Then the evolution in the rotating frame is

$$\mathbf{R}^{-1}\dot{\mathbf{s}}^r = -\frac{1}{2}\left(\mathbf{\Omega}_t \times \mathbf{s} - \mathbf{s} \times \mathbf{\Omega}_t\right)$$
$$-\frac{1}{2}\int_0^t \frac{((\mathbf{G}(t-t')\cdot\mathbf{\Omega}\times\mathbf{R}(t')\cdot\mathbf{s}(t'))\times\mathbf{s}(t)}{-\mathbf{s}(t)\times(\mathbf{G}(t-t')\cdot\mathbf{\Omega}\times\mathbf{R}(t')\cdot\mathbf{s}(t')))}dt'$$ (A.20)

Where we neglect the terms with $\dot{\mathbf{s}}^r$ compared with $\mathbf{\Omega} \times \mathbf{R} \cdot \mathbf{s}$ in the integral ("rotating frame approximation") [11]. We apply the rule that the vector product does not change if all the vectors experience the same rotation

$$\mathbf{F} = \mathbf{A} \times \mathbf{B}, \text{ then } \mathbf{R}\mathbf{F} = \mathbf{R}\mathbf{A} \times \mathbf{R}\mathbf{B}.$$ (A.21)

We obtain the desired evolution equation in the rotating frame

$$\dot{\mathbf{s}}^r = -\frac{1}{2}\left(\mathbf{\Omega}_t^r \times \mathbf{s}^r - \mathbf{s}^r \times \mathbf{\Omega}_t^r\right)$$
$$-\frac{1}{2}\int_0^t \frac{((\mathbf{G}^r(t-t')\cdot\mathbf{\Omega}\times\mathbf{s}^r(t'))\times\mathbf{s}^r(t)}{-\mathbf{s}^r(t)\times(\mathbf{G}^r(t-t')\cdot\mathbf{\Omega}\times\mathbf{s}^r(t')))}dt'$$ (A.22)

Now we make a Markov approximation [10] amounting to the assumption that the correlation time of the reservoir, i.e. the characteristic time over which functions in (A.4) and (A.5) change, is shorter than the coherent evolution time. Then the contribution of the reservoir can be approximated by the first terms of the integral series

$$
\mathbf{s}^r(t) - \mathbf{s}^r(0) =
$$

$$
-\frac{1}{2} \int_0^t dt' \left( \mathbf{\Omega}_t^r(t') \times \mathbf{s}^r(0) - \mathbf{s}^r(0) \times \mathbf{\Omega}_t^r(t') \right)
$$

$$
+\frac{1}{4} \int_0^t dt' \int_0^{t'} dt'' \begin{pmatrix} \mathbf{\Omega}_t^r(t') \times (\mathbf{\Omega}_t^r(t'') \times \mathbf{s}^r(t'')) \\ -(\mathbf{\Omega}_t^r(t'') \times \mathbf{s}^r(t'')) \times \mathbf{\Omega}_t^r(t') \end{pmatrix} \cdot
$$

$$
-\frac{t}{2} \int_0^{t'} ((\mathbf{G}^r(t-t') \cdot \mathbf{\Omega} \times \mathbf{s}^r(t')) \times \mathbf{s}^r(t) \\ -\mathbf{s}^r(t) \times (\mathbf{G}^r(t-t') \cdot \mathbf{\Omega} \times \mathbf{s}^r(t'))) dt'
$$

(A.23)

Further, averaging over the randomly states of the reservoir (designated by $\langle ... \rangle$) has to be performed. We take the reservoir to be in thermal equilibrium so that

$$
\langle a_n \rangle = \langle a_n^+ \rangle = \langle a_n^+ a_n^+ \rangle = \langle a_n a_n \rangle = 0 .
$$

(A.24)

$$
\langle a_n^+ a_n \rangle = n(\omega_n), \ \langle a_n a_n^+ \rangle = n(\omega_n) + 1 .
$$

(A.25)

The occupation number of the oscillator mode is such that

$$
2n(\omega) + 1 = \coth\left( \frac{\hbar\omega}{2k_B T} \right) .
$$

(A.26)

The averaging over the first power of a fluctuating variable gives zero $\langle \mathbf{\Omega}_t \rangle = 0$ . The matrix of averages of products of the reservoir variables is called the correlation function.

$$
\mathbf{C}_\Omega(t) = \langle \mathbf{\Omega}_t(0) \otimes \mathbf{\Omega}_t(t) \rangle .
$$

(A.27)

The Fourier transform of the correlation function is the spectral density of the reservoir

$$
\mathbf{S}_\Omega(\omega) = \int_{-\infty}^{\infty} dt e^{-i\omega t} \mathbf{C}_\Omega(t) .
$$

(A.28)

Direct substitution of (A.9) expresses it via the coupling constants (the negative frequencies in (A.28) are mapped on the positive frequencies of the reservoir)

$$
\mathbf{S}_\Omega(\omega) = \sum_n |g_n|^2 \left( 2n(\omega_n) + 1 \right) 2\pi\delta\left( \omega - \omega_n \right) \mathbf{n} \otimes \mathbf{n} .
$$

(A.29)

We also see that it is a symmetric tensor. The dissipation tensor $\mathbf{\Gamma}(\omega)$ is defined such that

$$\mathbf{S}_{\Omega}(\omega) = \mathbf{\Gamma}(\omega)\omega\big(2n(\omega)+1\big)/\hbar \, .$$

(A.30)

By substitution of (A.11) and comparing with (A.29), one concludes that

$$\int_{-\infty}^{t} \mathbf{G}(t-t')dt' = \mathbf{\Gamma}(0) \, .$$

(A.31)

The Markov approximation also enables one to extend the limits of integration of these functions to infinity. The system variables can be approximated as constants and taken out of the integral. With all the above equalities, the averaged evolution equation becomes

$$\begin{aligned}
\dot{\mathbf{s}}^{r} &= \frac{1}{2}\Big(\mathbf{S}_{\Omega}{}^{r}(0)\mathbf{s}^{r} - \mathbf{s}^{r}Tr\big(\mathbf{S}_{\Omega}{}^{r}(0)\big)\Big) \\
&\quad -\frac{1}{2}(\mathbf{\Gamma}^{r}(0)(\mathbf{\Omega}\times\mathbf{s}^{r}))\times\mathbf{s}^{r} + \frac{1}{2}\mathbf{s}^{r}\times(\mathbf{\Gamma}^{r}(0)(\mathbf{\Omega}\times\mathbf{s}^{r}))
\end{aligned}$$

(A.32)

From here on $\mathbf{s}$ means expectation value of the spin operator, averaged over the states of the reservoir. Simplifying the last term

$$\begin{aligned}
\dot{\mathbf{s}}^{r} &= \frac{1}{2}\Big(\mathbf{S}_{\Omega}{}^{r}(0)\mathbf{s}^{r} - \mathbf{s}^{r}Tr\big(\mathbf{S}_{\Omega}{}^{r}(0)\big)\Big) \\
&\quad +\left(\frac{\hbar}{2}\right)^{2}\big(\mathbf{\Omega}Tr(\mathbf{\Gamma}^{r}(0)) - \mathbf{\Omega}\mathbf{\Gamma}^{r}(0)\big)
\end{aligned}$$

(A.33)

Now we need to transform the equation back to the lab frame. By using (A.14) and (A.18), one can obtain the explicit form of the transform for the spectral density (if it is diagonal in the lab frame)

$$\mathbf{S}_{\Omega}{}'(\omega) = \frac{1}{2}\begin{pmatrix} S_{xx}(\omega+\Omega)+S_{xx}(\omega-\Omega) & iS_{xx}(\omega+\Omega)-iS_{xx}(\omega-\Omega) & 0 \\ iS_{xx}(\omega-\Omega)-iS_{xx}(\omega+\Omega) & S_{yy}(\omega+\Omega)+S_{yy}(\omega-\Omega) & 0 \\ 0 & 0 & 2S_{zz}(\omega) \end{pmatrix} .$$

(A.34)

The same form is true for the dissipation tensor. Then for coefficients of the evolution equation

$$\mathbf{S}_{\Omega}{}^{r}(0) = \begin{pmatrix} S_{xx}(\Omega) & 0 & 0 \\ 0 & S_{yy}(\Omega) & 0 \\ 0 & 0 & S_{zz}(0) \end{pmatrix} .$$

(A.35)

Substituting to (A.32), and transforming back to the lab frame one obtains the Bloch equations (we also take the spectral density to be isotropic for simplicity)

$$\dot{\mathbf{s}} = -\mathbf{\Omega}\times\mathbf{s} - \boldsymbol{\gamma}\big(\mathbf{s}-\mathbf{s}_{eq}\big) \, .$$

(A.36)

where the relaxation matrix

$$\gamma = \begin{pmatrix} \gamma_t & 0 & 0 \\ 0 & \gamma_t & 0 \\ 0 & 0 & \gamma_l \end{pmatrix}, \tag{A.37}$$

thermodynamic equilibrium value of spin in the external field is

$$\mathbf{s}_{eq} = \frac{\hbar}{2} \tanh\left(\frac{\hbar\Omega}{2k_B T}\right) \hat{\mathbf{z}}, \tag{A.38}$$

and the longitudinal and transverse relaxation rates are

$$\gamma_l(\Omega) \equiv \frac{1}{T_1(\Omega)} = S_\Omega(\Omega), \tag{A.39}$$

$$\gamma_t(\Omega) \equiv \frac{1}{T_2(\Omega)} = \frac{1}{2}\left(S_\Omega(\Omega) + S_\Omega(0)\right). \tag{A.40}$$

They are otherwise expressed via the longitudinal a transverse relaxation times, $T_1, T_2$, respectively.

We emphasize that the relaxation rates strongly depend on the spin splitting (given by $\Omega$). This dependence is one of the most important inputs used in this article.

## 10    APPENDIX B. SPIN RELAXATION I NQUANTUM DOTS

Now we turn to specific mechanisms of spin relaxation and obtain numerical estimates of for the case of GaAs quantum dots in order to illustrate the case of non-equilibrium spintronic computing.

Spins always coupled to the reservoir of the electromagnetic field modes, which is described by the Hamiltonian (3.2). The form of this coupling is universal. The magnetic field of the reservoir is [10] (see Appendix A for definitions of creation and annihilation operators and modes)

$$\mathbf{B} = \sum_n \sqrt{\frac{\hbar\omega_n\mu_0}{2V}} \left(a_n e^{-i\phi} + a_n^+ e^{i\phi}\right) \mathbf{n}, \tag{B.1}$$

where $V$ is the quantization volume, $\mu_0$ is the diamagnetic constant, $\omega_n$ is the frequency of the mode, and $\phi$ is the arbitrary phase. Using the results of Appendix A, we substitute the sum over the reservoir modes by the integral over frequency,

$$\sum_n \rightarrow \int \mathcal{D}(\omega) d\omega, \tag{B.2}$$

where the density of states is

$$\mathcal{D}(\omega) = \frac{V\omega^2}{\pi^2 c^3}. \tag{B.3}$$

Then by (A.29) and (A.39) the longitudinal relaxation rate is

$$\gamma_l(\Omega) = \frac{\hbar\Omega^3 G^2 \mu_0 (2n(\Omega)+1)}{\pi c^3}. \tag{B.4}$$

Numerical estimates show that the associated relaxation rate is too slow ($\gg 1/s$, see Figure 5) to be of practical significance. In a solid state, spin relaxation due to phonon coupling dominates. The physics of this coupling strongly depend on the properties of the material. Relaxation of spins localized in quantum dots, mostly in application to GaAs, has been considered in [12]. We use the material parameters of the GaAs substrate for the quantum dots. We also assume that the g-factor of the electron can be engineered to have a value up to

$g = 20$ by choosing the composition of the quantum dot. The are described by the Hamiltonian similar in form to the spin-photon Hamiltonian (B.1)

$$H_{spin-phonon} = \sum_n \frac{1}{2}\left\{ g_n^* a_n + g_n a_n^+, \mathbf{s}\cdot\mathbf{n} \right\}, \tag{B.5}$$

The details of the coupling constants $g_n$ and the summation over modes in this case are more involved than (B.1)-(B.4), and we refer the reader to [12] for details. Here we only reproduce the results for relaxation rates.

Spin relaxation happens in quantum dots due to spin-orbital interaction. Intuitively, the mechanism of relaxation is as follows: phonons induce local strain, corresponding change in the local electric field ; this changes the orbital motion of the electrons in the quantum dot; this in turn changes the spin of the electrons. There are two contributions to the spin-phonon Hamiltonian [12]

$$H_{spin-phonon} = H_{so} + H_{str}. \tag{B.6}$$

The first is the Dresselhaus spin-orbital interaction in materials with bulk-inversion asymmetry [12]

$$H_{so} = \beta(\sigma_y p_y - \sigma_x p_x). \tag{B.7}$$

where $p$ is the momentum operator of an electron, $\sigma$ are the Pauli matrices, and the coupling constant [12]

$\beta \sim 2\cdot10^3\, eV/m$ in GaAs.

This interaction causes admixture to the state with one projection of spin of another projection of spin. The corresponding relaxation rate in a dot with binding energy $\hbar\omega_0 \sim 0.5 eV$ is derived in [12] to be

$$\gamma_{l,adm}(\Omega) = \frac{\Omega^5}{\omega_0^4} \frac{4\left(eh_{14}\right)^2 \beta^2}{35\pi\rho\hbar}\left(\frac{1}{s_l^2} + \frac{4}{3s_t^2}\right)(2n(\Omega)+1), \tag{B.8}$$

where an element of piezotensor $eh_{14} \sim 1.2\cdot10^5\, eV/m$, density $\rho \approx 5\cdot10^3\, kg/m^3$, $s_l, s_t$ are longitudinal and transverse sound velocities [12].

The second is the direct coupling of spin to the strain field of the phonons [12]

$$H_{str} = \frac{1}{2}V_0\boldsymbol{\sigma}\cdot\varphi,$$ (B.9)

Where [12] $V_0 \approx 8\cdot10^5\,m/s$, and $\varphi$ are anticommutators of the lattice strain tensors with the electron momentum operators. The corresponding relaxation rate [12] is

$$\gamma_{l,dir}(\Omega) = \frac{\hbar\Omega^5 V_0^2}{240\pi\rho s^7}\frac{\Omega}{\sqrt{\Omega^2+\omega_0^2}}\big(2n(\Omega)+1\big),$$ (B.10)

Since the relaxation rates (B.8) and (B.10) are caused by phonons, these expressions are only valid for the frequency less than a certain cutoff frequency $\Omega < \omega_c$. This cutoff frequency is related to the maximum energy of the phonons and in turn depends on the material and the structure of quantum dots. In this calculation we will take it to be $\hbar\omega_c = 75meV$.

These two spin-phonon relaxation rates are shown in Figure 5. It shows that the relaxation due to direct coupling always exceeds the one due to spin-orbit coupling. However with the different set of material and quantum dot parameters, it may be reversed [12]. Both these mechanism have a very steep, fifth power dependence on the spin splitting energy and turn to zero at zero splitting.

Different mechanisms must be dominant at zero splitting. One of them is the magnetic noise – interaction with the magnetic field generated by the fluctuating currents in conductors around the dot (mostly the metal wires). The spectral density of the current fluctuations in a wire (Johnson noise)

$$S_I(\omega) = \frac{\hbar\omega}{R}\big(2n(\omega)+1\big),$$ (B.11)

is determined [11] by the resistance of the wire $R = l/(\sigma_c w^2)$, where $l$ is the length of the wire, $w$ is the cross-section size, and $\sigma_c$ is the conductivity of the material.

We apply the Bio-Savar law for the magnetic field of a current loop

$$B = \frac{\mu_0 I}{2\pi r},$$ (B.12)

and (3.2) to determine the spectral density of the reservoir $S_\Omega(\omega)$ according to (A.28) and hence the longitudinal relaxation according to (A.39)

$$\gamma_{l,res}(\Omega) = \frac{\hbar\Omega G^2 \mu_0^2 \sigma_c w^2}{4\pi^2 r^2 l}\big(2n(\Omega)+1\big),$$  (B.13)

The rate is estimated for the wire length $l = 300nm$ and distance from the dot to the wire $r \sim 10nm$, the wire cross-section size of $w \sim 10nm$, for the conductivity of copper $\sigma_c = 6 \cdot 10^7 /(\Omega * m)$, and plotted in Figure 5 as well.

## 11  CORRELATION FUNCTION AND MAGNITUDE OF NOISE MAGNETIC FIELD

It is instructive to calculate the magnitude of the stochastic magnetic field and the time that it takes for the phase to change significantly (the correlation time).

The correlation function (A.27) gives the variance of the stochastic angular velocity

$$\left\langle \mathbf{\Omega}_t^{\,2}(0) \right\rangle = \mathbf{C}_\Omega(t=0) = \frac{1}{2\pi}\int_{-\infty}^{\infty} \mathbf{S}_\Omega(\omega)d\omega.$$  (C.1)

The corresponding r.m.s. value of the magnetic field is then obtained via the ratio (3.20)

$$\left\langle \mathbf{\Omega}_t^{\,2} \right\rangle = G^2 B_{rms}^{\,2}.$$  (C.2)

This translates to the conversion factor

$$\Omega / B = 1.77 \cdot 10^{12} /(s \cdot \text{Tesla}).$$  (C.3)

for $g = 200$. The spectra of noise in (B.8) and (B.10) are well approximated by the functional form

$$S_\Omega(\Omega) = A\Omega^\alpha, \quad \text{for } \Omega < \omega_c,$$  (C.4)

where $A$ is a constant and $\alpha = 5$. Integration in (C.1) yields

$$\left\langle \mathbf{\Omega}_t^{\,2} \right\rangle = \frac{1}{\pi}\int_0^{\omega_c} A\omega^\alpha d\omega = \frac{\gamma(\omega_c)\omega_c}{\pi(\alpha+1)}.$$  (C.5)

For the values in Table I, this results in

$$\sqrt{\left\langle \mathbf{\Omega}_t^{\,2} \right\rangle} = 2.4 \cdot 10^{14} / s, \quad B_{rms} \approx 14\text{T}$$  (C.6)

It seems like a very large magnetic field, but it represents a sum of rapidly fluctuating harmonics over a wide range of frequencies. It characterizes the fast relaxation at high spin splitting, such as $\Omega_c$. For smaller spin splitting, more relevant parameter is the noise magnetic field which is resonant to the transition, i.e. within the range $\gamma_l(\Omega)$ from the splitting frequency $\Omega$:

$$\left\langle \mathbf{\Omega}_t^{\,2} \right\rangle_{res} = \frac{1}{\pi} \int\limits_{\Omega-\gamma_l}^{\Omega+\gamma_l} \mathbf{S}_\Omega(\omega) d\omega \approx \frac{2\gamma_l^{\,2}}{\pi}. \tag{C.7}$$

This results in the field of $B_{rms} \approx 1.4 \cdot 10^{-7}\,\mathrm{T}$ at $\Omega_0$. This is consistent with the observation that the relaxation is much slower than precession at the computing frequency and much faster than precession at the clocking frequency.

In order to estimate how quickly the memory of the phase of the fluctuating magnetic field is lost, one needs to determine the time dependence of the correlation function (A.27). It can be derived from the spectrum of noise (A.28). Using the fact that the spectrum is the same for negative and positive frequencies

$$\mathbf{C}(t) = \frac{1}{\pi} \int\limits_0^\infty \mathbf{S}_\Omega(\omega) \cos(\omega t) d\omega, \tag{C.8}$$

Expanding in a Taylor series for small value of t

$$\mathbf{C}(t) = \frac{1}{\pi} \int\limits_0^{\omega_c} A\omega^\alpha \left( 1 - \frac{(\omega t)^2}{2} + \dots \right) d\omega,$$

$$= \frac{\gamma(\omega_c)\omega_c}{\pi(\alpha+1)} - \frac{\gamma(\omega_c)\omega_c^{\,3} t^2}{2\pi(\alpha+3)} + \dots \tag{C.9}$$

one concludes that the correlation function significantly decreases at the time interval ("correlation time").

$$\tau_{corr} = \frac{1}{\omega_c} \sqrt{\frac{2(\alpha+3)}{(\alpha+1)}}. \tag{C.10}$$

Though the reservoir is in thermal equilibrium, the above derivation shows that the correlation time is not determined by the temperature.

REFERENCES

[1] Semiconductor Industry Association, "International Roadmap for Semiconductors", chapter "Emerging Research Devices" (2003). Available: http://public.itrs.net/Files/2003ITRS/Home2003.htm.

[2] V. V. Zhirnov, R. K. Cavin, J. A. Hutchby, and G. I. Bourianoff, "Limits to binary logic switch scaling – a gedanken model", Proceedings of IEEE, 91, 1934 (2003).

[3] R. Landauer, "Irreversibility and heat generation in the computing process", IBM Journal of Research and Development, 5, 183 (1961).

[4] S. A. Wolf et al., "Spintronics: a spin-based electronics vision for the future", Science, 294, 1488 (2001).

[5] I. Zutic, J. Fabian, and S. Das Sarma, "Spintronics: fundamentals and applications", Reviews of Modern Physics, 76, 323 (2004).

[6] R. P. Feynman, Feynman Lectures on Computation, (Addison Wesley, Reading, MA, 1996).

[7] K. K. Likharev, "Dynamics of some single flux quantum devices. I. Parametric quantron.", IEEE Trans. Magn., 25, 1436 (1989).

[8] S. Bandyopadhyay, "When it comes to spintronics, there may be some room in the middle", cond-mat 0412519, (2004). available at http://www.arxiv.org/.

[9] J. Timler and C. S. Lent, "Maxwell demon and quantum dot cellular automata", J. Appl. Phys., 94, 1050 (2003).

[10] C. W. Gardiner and P. Zoller, Quantum Noise: A Handbook of Markovian and Non-Markovian Quantum Stochastic Methods With Applications to Quantum Optics, (Springer, 2nd edition, 1999), Chapter 3.

[11] J. A. Sidles, J. L. Garbini, W. M. Dougherty, and S.-H. Chao, "The classical and quantum theory of thermal magnetic noise, with applications in spintronics and quantum microscopy", Proceedings of IEEE, 91, 799 (2003).

[12] A. V. Khaetskii and Y. V. Nazarov, "Spin-flip transitions between Zeeman sublevels in semiconductor quantum dots", Physical Review B, 64, 125316 (2001).

[13] G. Csaba, A. Imre, G. H. Bernstein, W. Porod, and V. Metlushko, "Nanocomputing by field-coupled nanomagnets", IEEE Tran. On Nanotechnology, 1, 209 (2002).

[14] S. Bandyopadhyay, B. Das, and A. E. Miller, "Supercomputing with spin-polarized single electrons in a quantum coupled architecture", Nanotechnology, 5, 113 (1994).

[15] M. C. B. Parish and M. Forshaw, "Physical constraints on magnetic quantum cellular automata", Appl. Phys. Lett., 83, 2046, (2003).

[16] N. Margolus and L. B Levitin, "The maximum speed of dynamical evolution", Physica D, 120, 188 (1998).

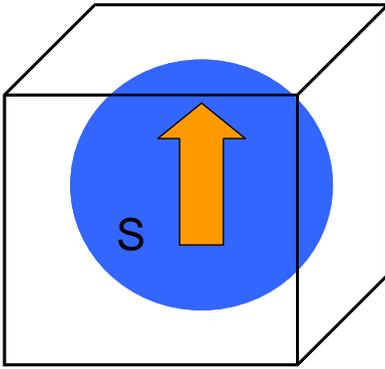

Fig. 1. Single electron with spin confined to a quantum dot.

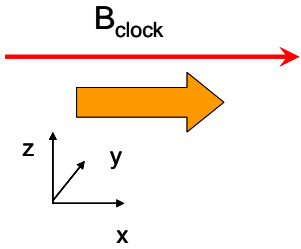

Fig. 2a. Spin state in computing, a) clocking.

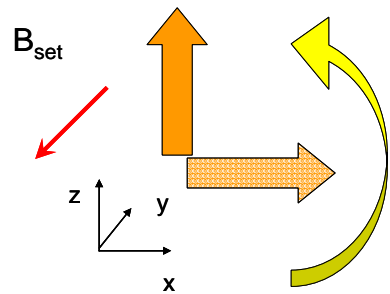

Fig. 2b. Spin state in computing, b) setting the initial value.

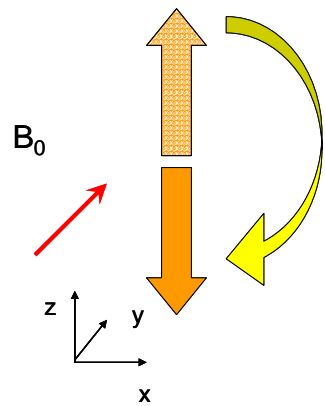

Fig. 2c. Spin state in computing, c) switching.

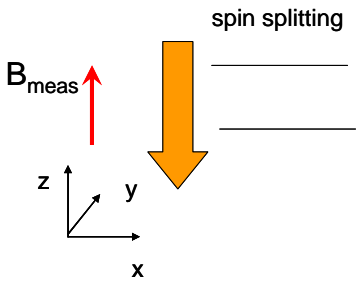

Fig. 2d. Spin state in computing, d) read-out.

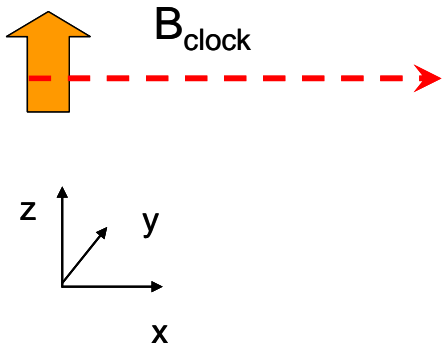

Fig. 2e. Spin state in computing, e) before the next clocking.

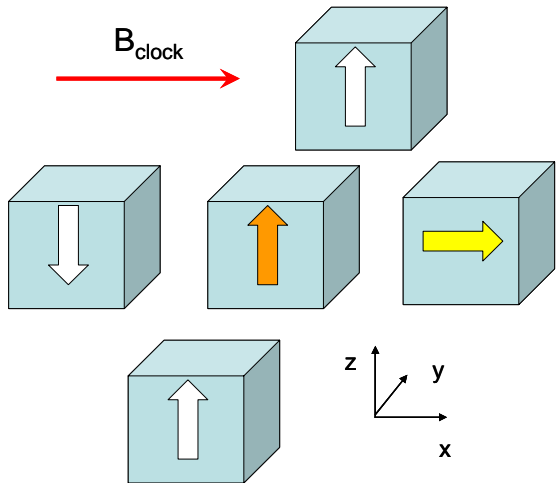

Fig. 3. Spintronic circuit with multiple quantum dots; a majority gate.

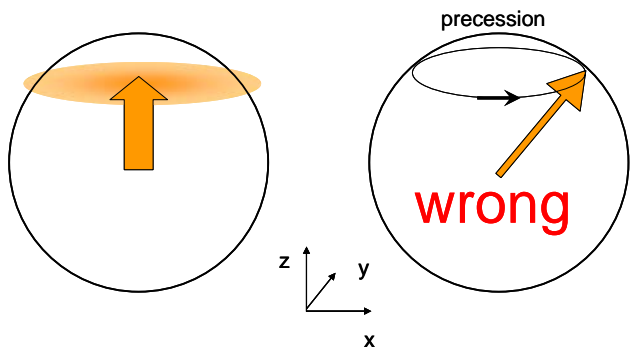

Fig. 4. Projections and variances of the spin (left). It is not correct to explain the quantum uncertainty as precession with random phase around the expectation value of the projection.

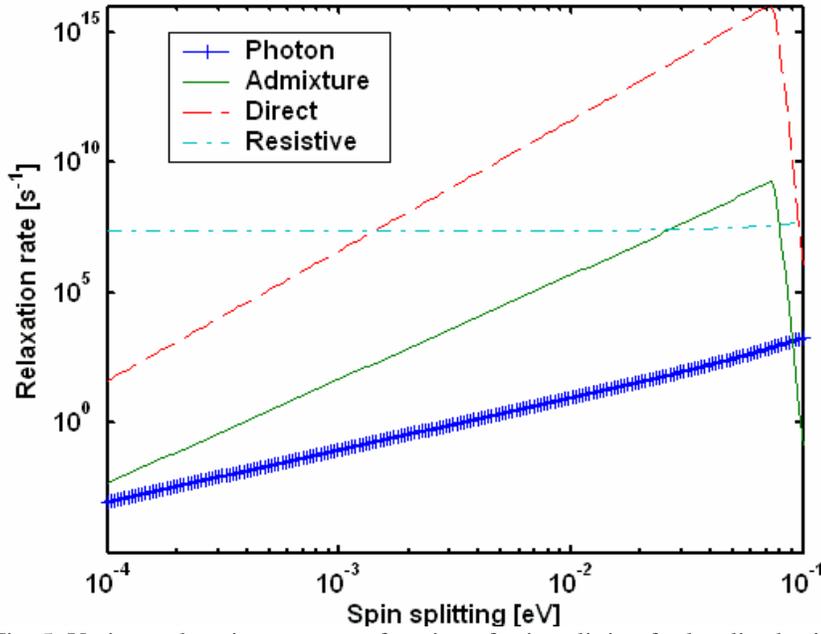

Fig. 5. Various relaxation rates as a function of spin splitting for localized spins in quantum dots: circles – due to coupling to the electromagnetic field (photons), solid line – due to phonon coupling with admixture of excited states, dashed line- due to direct spin coupling to phonon-generated strain, dash-dotted line – due to fluctuations of magnetic field caused by current.

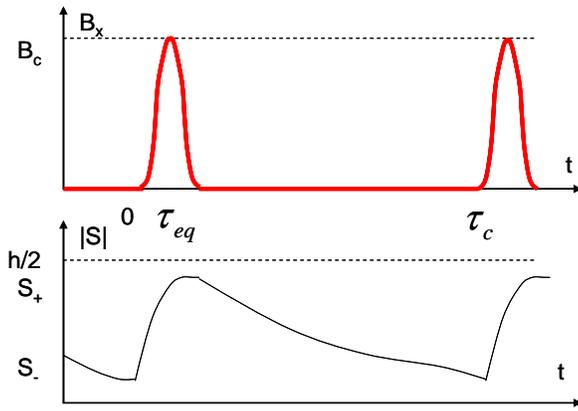

Fig. 6. Time dependence of the clocking field and the spin magnitude over the clocking cycle.

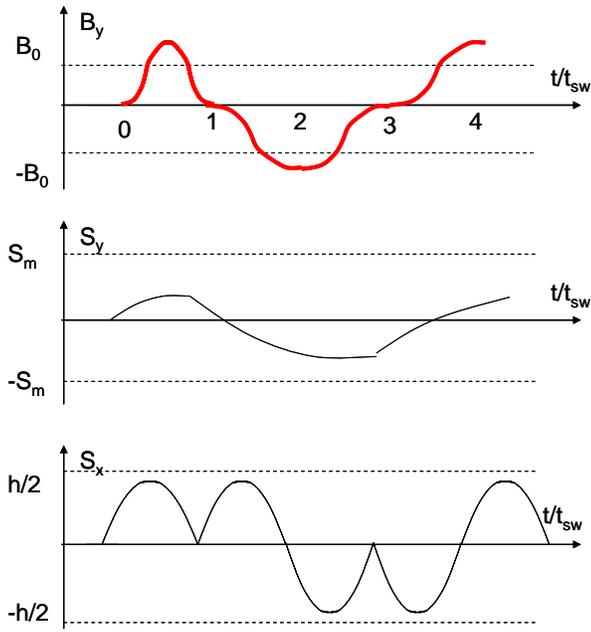

Fig. 7. Time dependence of a longitudinal and a transverse projections of spin as the switching field is varied. A few switching cycles are shown.